\documentclass[prd,preprintnumbers,twocolumn,eqsecnum,floatfix,nofootinbib,a4paper,superscriptaddress]{revtex4-1}
\usepackage{color}
\usepackage{calc}
\usepackage{amsmath,amssymb,graphicx}
\usepackage{amssymb,amsmath}
\usepackage{tensor}
\usepackage{bm}
\usepackage{float}
\usepackage{microtype}
\usepackage{booktabs}
\usepackage{times}
\usepackage[varg]{txfonts}
\usepackage[colorlinks, pdfborder={0 0 0}]{hyperref}
\usepackage{subfigure}
\definecolor{LinkColor}{rgb}{0.75, 0, 0}
\definecolor{CiteColor}{rgb}{0, 0.5, 0.5}
\definecolor{UrlColor}{rgb}{0, 0, 0.75}
\hypersetup{linkcolor=LinkColor}
\hypersetup{citecolor=CiteColor}
\hypersetup{urlcolor=UrlColor}
\allowdisplaybreaks
\usepackage[utf8]{inputenc}
\usepackage{ulem}
\usepackage{comment}
\usepackage{hyperref}
\usepackage[nolist]{acronym}
\usepackage{tikz}
\usepackage{textcomp}
\usepackage[export]{adjustbox}
\usepackage{url}

\usetikzlibrary{arrows,shapes,trees,decorations.pathreplacing}
\usepackage{aas_macros}

\allowdisplaybreaks

\begin{document}

\title{Rapid Identification of Strongly Lensed Gravitational-Wave Events with Machine Learning}

\author{Srashti Goyal}
\affiliation{International Centre for Theoretical Science, Tata Institute of Fundamental Research, Bangalore - 560089, India}

\author{Harikrishnan D.}
\affiliation{International Centre for Theoretical Science, Tata Institute of Fundamental Research, Bangalore - 560089, India}
\affiliation{Plaksha Tech Leaders Fellowship}

\author{Shasvath J. Kapadia}
\affiliation{International Centre for Theoretical Science, Tata Institute of Fundamental Research, Bangalore - 560089, India}

\author{Parameswaran Ajith}
\affiliation{International Centre for Theoretical Science, Tata Institute of Fundamental Research, Bangalore - 560089, India}
\affiliation{ Canadian Institute for Advanced Research, CIFAR Azrieli Global Scholar, MaRS Centre, West Tower, 661 University Ave, Toronto, ON M5G 1M1, Canada}


\begin{abstract}
A small fraction of the gravitational-wave (GW) signals that will be detected by second and third generation detectors are expected to be strongly lensed by galaxies and clusters, producing multiple observable copies. While optimal Bayesian model selection methods are developed to identify lensed signals, processing tens of  thousands (billions) of possible pairs of events detected with second (third) generation detectors is both computationally intensive and time consuming. To  mitigate this problem, we propose to use machine learning to rapidly rule out a vast majority of candidate lensed pairs. As a proof of principle, we simulate non-spinning binary black hole events added to Gaussian noise, and train the machine on their  time-frequency maps (Q-transforms) and localisation skymaps (using Bayestar), both of which can be generated in seconds.  We show that the trained machine is able to accurately identify lensed pairs with efficiencies comparable to existing Bayesian methods. 
\end{abstract}

\maketitle

\section{Introduction}\label{sec:introduction}
	With the tens of gravitational-wave (GW) signals detected by the LIGO-Virgo network of detectors \cite{aLIGO, Virgo} during its first three observing runs \cite{gwtc-2, gwtc-1, ias-1, ias-2, ias-3, ias-4}, there is no doubt that GW astronomy has well and truly arrived. While these detections have enabled stringent tests of Einstein's general relativity \cite{tgr-gwtc1, tgr-gwtc2}, future observing runs are likely to provide a number of additional tests. Among them is the highly anticipated observation of gravitationally-lensed GWs \cite{ohanian1974, bliokh1975, bontz1981, Deguchi1986, Nakamura1998}, akin to the gravitational lensing of electromagnetic waves where the deflection of light from a source due to large agglomerations of matter (such as galaxies and galaxy clusters) residing along the line of sight of the observer produces multiple magnified (or demagnified) copies of the source. Apart from being a unique probe of general relativity's prediction of gravitational lensing with a different messenger \cite{Fan2017}, lensing of GWs could afford unique constraints on various aspects of astrophysics and cosmology, including models of the populations of galaxies \cite{Smith2019}, as well as models that probe the distribution and composition of dark matter \cite{Jung2019}. 

Gravitational lensing observations are typically divided into three categories: strong lensing, weak lensing and micro lensing (see, e.g., \cite{dodelson2017}). This classification is based on the properties of the lens, in particular the density of the lens projected along the plane perpendicular to the line of sight of the observer. In this work, we concern ourselves exclusively with strong lensing, where the projected density exceeds a critical density, resulting in the production of multiple resolvable images. 


Note that this classification is done in the \emph{geometric optics} limit, where the wavelength (of light or GWs) is much smaller than the characteristic gravitational radius of the lens. While this is almost always true in the lensing of light, this is not always the case for the lensing of GWs. In this work, we assume that the wavelength of the GWs is much smaller than the Schwarzchild radius of the lenses, as is the case when GWs from coalescing stellar-mass binary black holes are lensed by galaxies or galaxy clusters. In this  limit, strongly lensed GWs will result in the production of potentially resolvable images.

The resolvability of images in the sky is ultimately dependent on the resolution of the telescopes that observe these images. GW detectors typically have very poor angular resolution \cite{fairhurst2009, fairhurst2018} (at least in comparison to optical telescopes); the localisation skyarea for GW events detected by the LIGO-Virgo network in the second and third observing runs spanned tens of square degrees at best \cite{gwtc-2}. As a result, even strongly lensed gravitational-wave events typically have images whose skyareas almost completely overlap each other. Indeed, one of the signatures that two GW events are lensed copies is that their skymaps overlap (see, e.g, \cite{haris2018, Singer2019}).  

While strongly lensed GW events are completely unresolvable in the sky with current GW detectors, they are typically very well resolved in time. Indeed, the temporal resolution of GW events (~miliseconds) is in general orders of magnitude smaller than the expected time delay (minutes to weeks) between strongly lensed GW images.  In the geometric optics limit, these GW images would have different amplitudes, but their phase evolution would be identical \cite{Deguchi1986, Wang1996, Nakamura1998, Takashi2003, Dai2017, Ezquiaga2020}. Thus, in principle, determining whether two non-overlapping GW events are lensed copies comes down to comparing the shapes of these signals with respect to each other.

In practice, however, such a comparison is non-trivial. Firstly, the observed GW signals are projections of the true GW signals onto the detectors; this projection depends on the location and orientation of the detector relative to the source, and would therefore be different for each of the temporally separated GW images. Furthermore, these images would be buried in detector noise. Even if the noise is assumed to be Gaussian and the corresponding power spectral density (PSD) is assumed to be time invariant, each of the images would be buried in different realisations of this noise. 

A robust alternative to such a direct comparison of the GW-signals is to work in the space of the inferred source parameters. Using optimal matched-filter based parameter inference techniques \cite{Veitch2015}, Bayesian posterior distributions on the intrinsic parameters of the source (the masses and spins of the binary) and its extrinsic parameters (the skylocation of the binary) can be constructed. As mentioned earlier, the phase evolution of the GW images are expected to be identical, and therefore comparing the inferred posteriors on the intrinsic parameters (which completely govern the phase evolution) of pairs of GW events should enable us to discriminate between lensed and unlensed pairs. This discriminatibility can be further enhanced by comparing the localisation skymaps which are expected to overlap almost entirely for lensed GW pairs \cite{haris2018}.

Quantitatively, such a comparison can be achieved using Bayesian model selction \cite{haris2018, hannuksela2019}. A Bayes factor derived from the overlap between the posteriors of pairs of events can be constructed and used to segregate these pairs as either lensed and unlensed. However, evaluating this discriminator is computationally expensive and time consuming. Bayesian parameter inference of BBH events can take hours to days. Additionally, constructing the Bayes factor can take up to a few minutes per event, and the number of such evaluations will grow as the square of the number GW events. This makes the estimation of the Bayes factor computationally challenging when large numbers of BBH events are expected to be detected in future observing runs.

Current estimates of the rate of stellar-mass binary black-hole (BBH) mergers \cite{o3apop} suggest that hundreds of BBH events are expected to be detected in LIGO-Virgo-Kagra's next observing run (O4). Among these GW detections, up to a percent could be lensed copies of each other \cite{Ng2018, Xu2021}, suggesting that there is a non-trivial chance that the first confirmed detection of a lensed GW pair could occur in O4. However, identifying such lensed pairs would require constructing $\mathcal{O}(10^{2})$ posteriors on the GW events' source-parameters and $\mathcal{O}(10^4)$ Bayes factors. 

These numbers will get significantly larger with observing runs beyond O4, and astronomically large by the time the third generation (3G) network of ground-based detectors \cite{CosmicExplorer, EinsteinTelescope, 3Gsensitivity} completes its observations. The 3G network is expected to observe $\mathcal{O}(10^5 - 10^6)$ events, of which $\sim 0.3\%$ could be strongly lensed \cite{Xu2021}. Therefore, $\mathcal{O}(10^5 - 10^6)$ event posteriors, and $\mathcal{O}(10^{10} - 10^{12})$ Bayes factors, would need to be evaluated.

This motivates the need to come up with a method to conduct a preliminary segregation of pairs of GW events to rapidly ``weed out'' the vast majority of unlensed pairs. In this work, we propose to use machine learning algorithms, trained on time-frequency maps of the detector strain time series \cite{chatterji2004} and the (rapidly estimated) localisation skymaps \cite{singer2016}, from both lensed and unlensed pairs of GW events, to construct a statistic to discriminate between lensed and unlensed pairs. Using synthetic, non-spinning BBH signals -  both lensed and unlensed - injected in Gaussian noise, we show that our machine-learning-based statistic, performs almost as well as the optimal Bayes factor statistic described above, while reducing the computation time by orders of magnitude. The significant reduction in evaluation time is a direct consequence of the fact that time-frequency maps and localisation skymaps can be constructed in seconds, in constrast to GW inference posteriors which take hours to days to sample.

The rest of this paper is organized as follows. Section~\ref{sec:method} summarizes the evaluation of the optimal Bayes factor statistic, introduces the machine learning algorithms we use, and delineates their training and validation. Section~\ref{sec:results} describes our results in distinguishing between lensed and unlensed GW event pairs and compares them with the performance of  the posterior overlap statistic. Section~\ref{sec:conclusion} summarizes this work and discusses its potential benefits.

\section{Method}\label{sec:method}
\subsection{The posterior overlap statistic}

Let $d(t)$ be the detector strain time series which is known to contain a gravitational wave signal $h(t, \vec{\theta)}$ with shape (intrinsic and extrinsic) parameters $\vec{\theta}$, as well as one realisation of stochastic Gaussian noise as characterized by its power spectral density $S_n(f)$. A Bayesian inference of $\vec{\theta}$ from $d(t)$ can be achieved by sampling the posterior distribution on $\vec{\theta}$:
\begin{equation}
p(\vec{\theta} \mid d) = \frac{p(\vec{\theta})p(d \mid \vec{\theta})}{p(d)}
\end{equation}
where \cite{cuttler-flanagan}:
\begin{equation}
p(d~|~\vec{\theta}) \propto \exp\left[-(d - h \mid d-h)/2\right]
\end{equation}
is the Gaussian likelihood, $p(\vec{\theta})$ is the prior distribution on the source parameters, $p(d)$ is the evidence, and $(\cdot\mid\cdot)$ symbolises the noise-weighted inner product:
\begin{equation}
(a~|~b)  \equiv 2\int_{f_{\mathrm{min}}}^{f_{\mathrm{max}}}\frac{\tilde{a}(f)\tilde{b}^{*}(f)}{S_n(f)}df
\end{equation}
Here, $\tilde{a}, \tilde{b}$ represent the Fourier transform of the time series $a(t), b(t)$; $[f_{\mathrm{min}}, f_{\mathrm{max}}]$ is the frequency range over which the inner product is evaluated; and $^*$ represents complex conjugation.

Now consider two segments of data, $d_1(t)$ and $d_2(t)$, both of which are known to contain one GW signal each, $h_1(t)$ and $h_2(t)$, respectively. We now wish to determine which of the two hypotheses, $\mathcal{H}_L$ and $\mathcal{H}_U$, is preferred by the data at hand. 

$\mathcal{H}_{L}$ is the hypothesis that $h_1(t)$ and $h_2(t)$ are lensed copies of a GW signal originating from a single source. On the other hand, $\mathcal{H}_{U}$ is the hypothesis that $h_1(t)$ and $h_2(t)$ are signals originating from two distinct, unrelated, sources.

As shown in \cite{haris2018}, (in the absence of any prior knowledge of which of the hypotheses is preferred), the optimal Bayesian statistic to quantitatively determine the preferred hypothesis is the Bayes factor $\mathcal{B}^L_U$, defined as the ratio of the evidences of the joint data set $\lbrace d_1, d_2 \rbrace$ given each of the hypotheses.
\begin{equation}
\mathcal{B}^L_U \equiv \frac{p(\lbrace d_1, d_2 \rbrace \mid \mathcal{H}_L)}{p(\lbrace d_1, d_2 \rbrace \mid \mathcal{H}_U)} = \int \frac{p(\vec{\theta} \mid d_1)p(\vec{\theta} \mid d_2)}{p(\vec{\theta})}d\vec{\theta}
\end{equation}
This Bayes factor can be evaluated making use of the posteriors $p(\vec{\theta}~|~d_1)$ and $p(\vec{\theta}~|~d_2)$ estimated from the two data sets $d_1$ and $d_2$, as well as the prior $p(\vec{\theta})$ employed in the parameter estimation.

\subsection{Classification with Machine Learning}

In the language of machine learning (ML), determining whether a pair of GW events are lensed copies of a single GW event, or unrelated (unlensed) to each other, is a binary classification problem. Using features derived from the data surrounding pairs of GW signals, we can in principle train an ML algorithm to classify them as either lensed or unlensed. In this subsection we first describe the construction of the features we use, the ML algorithms we employ, along with their training, testing and optimisation. 

\subsubsection{Data Representation}\label{sec:data_rep}

The posterior overlap statistic crucially relies on a time-consuming way of representing the detector data, viz., the posterior distributions of source parameters inferred from the data surrounding the confirmed GW detections. To bypass this issue, we construct and train a machine learning model which takes as inputs time-frequency maps ({\it Q-transforms} of the GW event), as well as localisation skymaps ({\it Bayestar Skymaps}). Both of these can be produced within seconds, in contrast to sampling the full posterior on the source parameters which can take anywhere from several hours to several days. 

{\it Q-Transforms}: Q-transforms \cite{chatterji2004} are a means by which time-frequency maps of generic transient signals can be produced. This is achieved by first representing the time-frequency plane as a collection of tiles (bins), and then reconstructing these generic signals as a combination of sine-Gaussians defined by their quality factor `Q'.  The choice of `Q' in each tile is determined from a matched-filter search across multiple `Q' templates, and the template that produces the largest SNR is selected. Using the corresponding optimal sine-Gaussian, a spectrogram is generated. The time-frequency map is then plotted as colored tiles, where the color represents the so-called ``normalized signal-energy'', which is proportional to the Q-transform magnitude (and related to the SNR). 

\begin{figure*}[htb]
\includegraphics[width=0.48\linewidth]{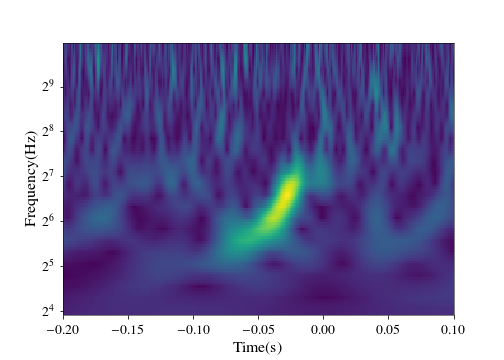}
\includegraphics[width=0.48\linewidth]{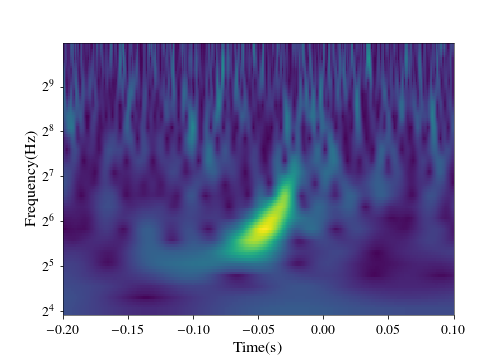}
\includegraphics[width=0.48\linewidth]{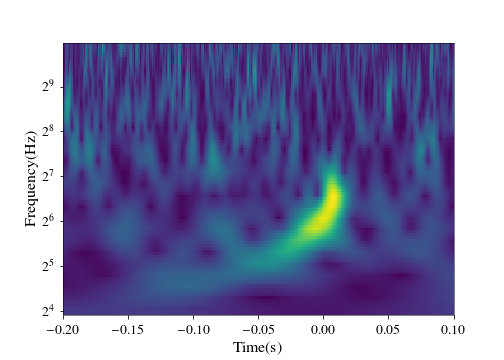}
\includegraphics[width=0.48\linewidth]{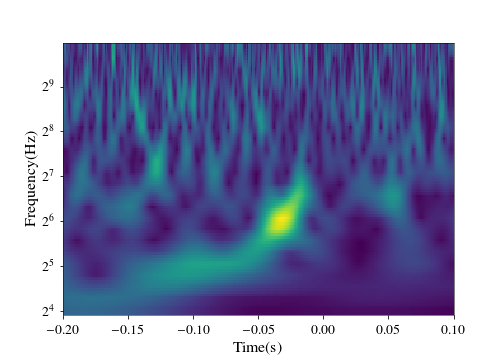}
\centering
\includegraphics[trim={0 0.3cm 0cm 7.5cm},clip,width=0.5\linewidth]{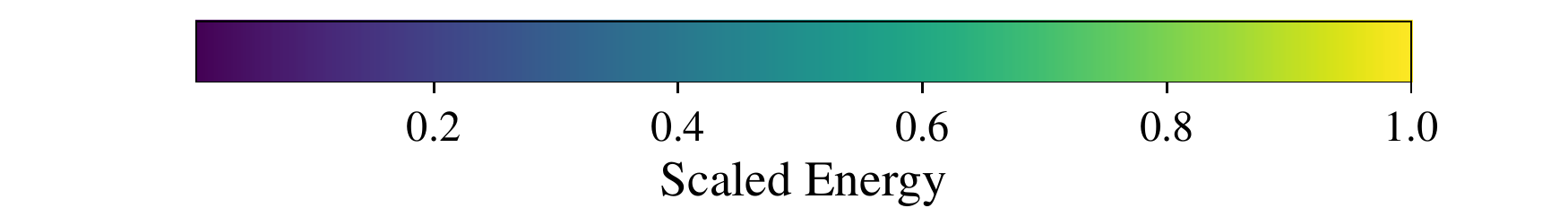}
\caption{{\it Top Panels}: A pair of lensed GW events detected by the H1 (Hanford) inteferometer at design sensitvity. These events have time-frequency tracks with similar shapes. However, the signal energy in different time frequency bins along their tracks differ with respect to each other. {\it Bottom Panels}: A pair of unlensed GW events projected detected by the H1 inteferometer at design sensitvity. These events have time-frequency tracks whose shapes are significantly different.}
\label{fig:qts}
\end{figure*}

As shown in Fig.~\ref{fig:qts}, lensed events will have time-frequency maps whose shapes are similar, but whose signal energies across time-frequency tiles will differ in magnitude. This is a direct consequence of the fact that the phase evolution of strongly lensed pairs are expected to be identical, but the amplitudes will differ by a constant factor. On the other hand, unlensed signals will have distinct time-frequency maps with dissimilar shapes in general. \footnote{A constant (additive) phase-factor called the Morse-phase, which is an integral multiple of $\pi/2$ depending on image type, will in general change the coalescence phase of the dominant GW mode \cite{Dai2017, Dai2020}. Note that Q-transforms are independent of coalescence phase, and are therefore unaffected by the Morse phase.}

{\it Bayestar Skymaps}: ``Bayestar'' \cite{singer2016} is the flagship low-latency skylocalisation software of the LIGO-Virgo-Kagra (LVK) collaboration, used during the LVK's third observing run (O3) to disseminate skymaps in real-time for electromagnetic follow-up of GW events \cite{o3lowlatency}. These skymaps are produced in seconds, and are found to be comparable to those estimated from a full sampling of the joint posterior distribution of the source parameters. Bayestar exploits the fact that errors in sky localisation and the errors in the inference of the source masses, are semi-independent. Given that this software is exclusively focussed on providing localisation skyareas, it exploits this semi-independence to drastically reduce the dimensionality of the parameter estimation problem by fixing the intrinsic parameter values to those of the maximum likelihood template in the matched filter search that identified the event. It is thus able to evaluate the (dimensionally-reduced) posterior on the extrinsic parameters rapidly, without significant loss in precision. 

\begin{figure*}
\includegraphics[trim={1.2cm 1.2cm 1.5cm 3cm},clip,width=0.48\linewidth]{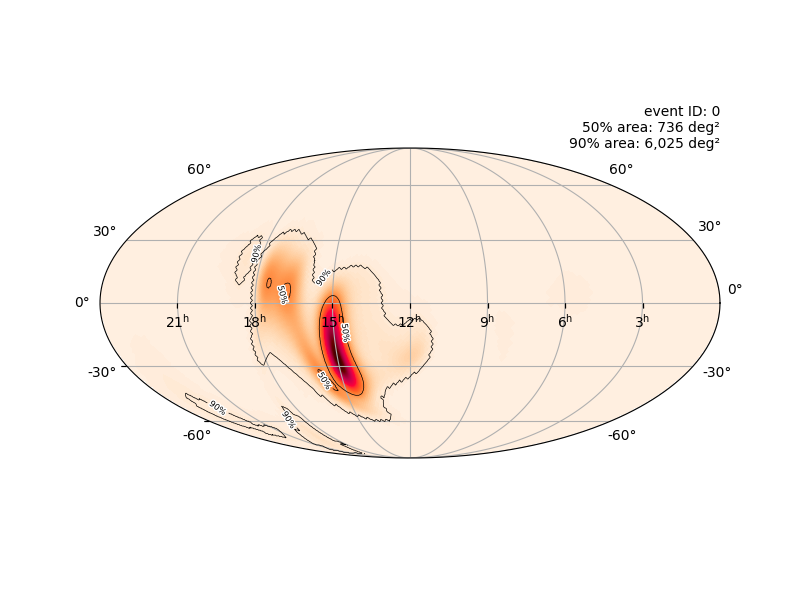}
\includegraphics[trim={1.2cm 1.2cm 1.5cm 3cm},clip,width=0.48\linewidth]{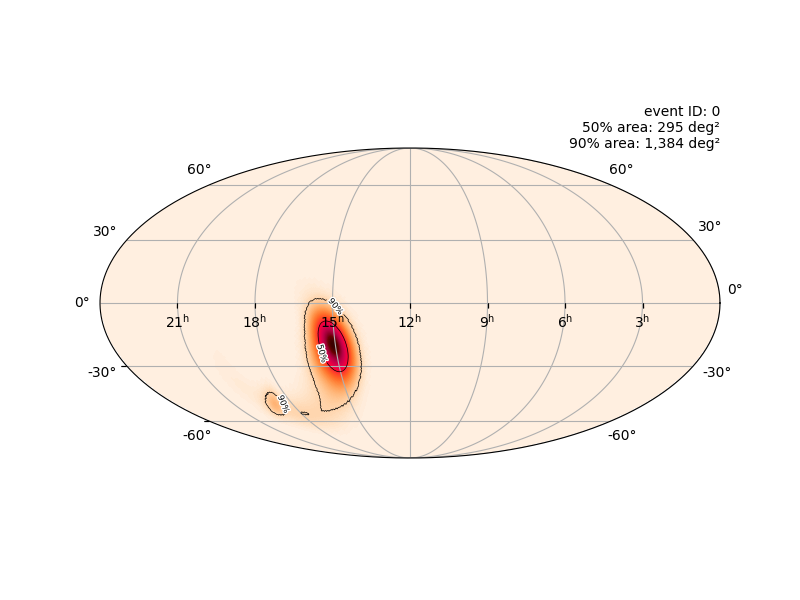}
\includegraphics[trim={1.2cm 1.2cm 1.5cm 3cm},clip,width=0.48\linewidth]{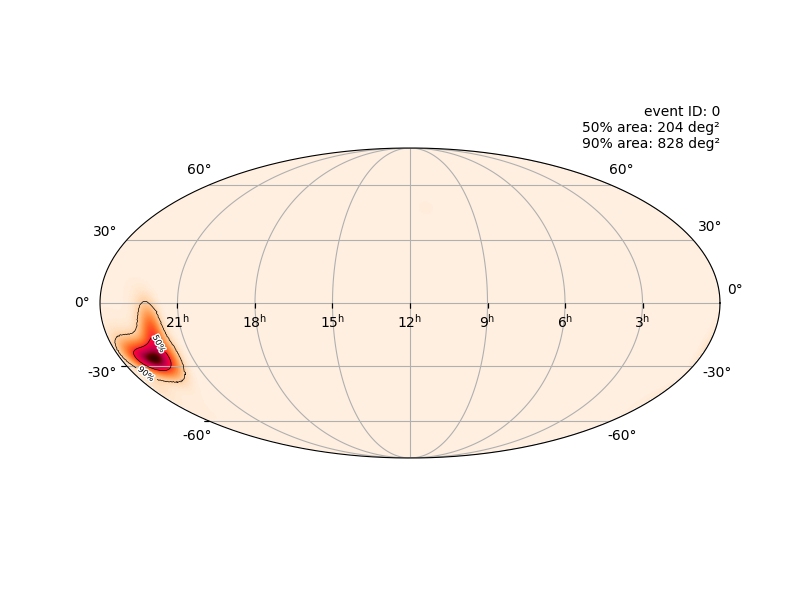}
\includegraphics[trim={1.2cm 1.2cm 1.5cm 3cm},clip,width=0.48\linewidth]{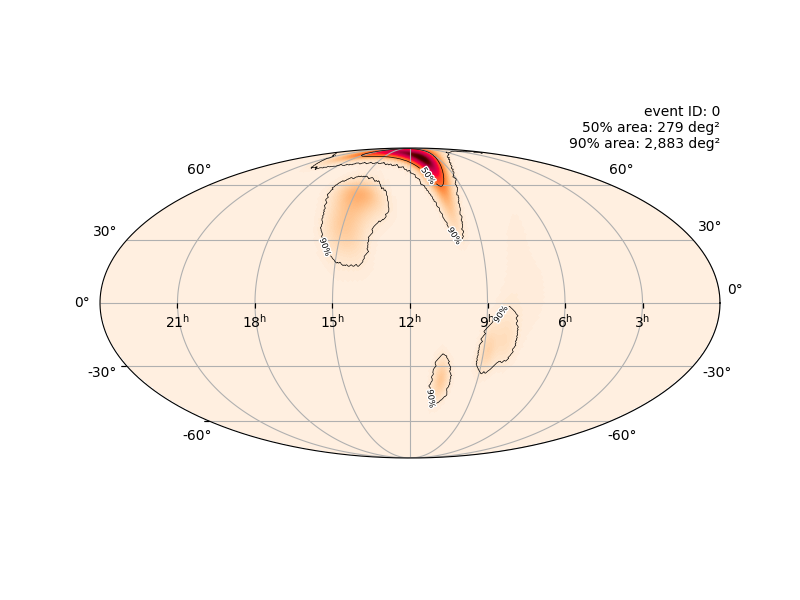}
\caption{{\it Top Panels}: Bayestar skymaps of a pair of lensed events detected by the H1 (Hanford), L1 (Livingston), V1 (Virgo) network at design sensitivity. The skymaps of these events overlap. {\it Bottom Panels}: Bayestar skymaps of a pair of unlensed events detected by the H1, L1, V1 network at design sensitivity. The skymaps of these events do not overlap.}
\label{fig:skymaps}
\end{figure*}

As shown in Fig.~\ref{fig:skymaps}, lensed events are expected to have overlapping localisation skyareas, by virtue of the poor ($\mathcal{O}(10)$ sq. deg.) angular resolution of ground based GW detectors with respect to the typical angular separation of the images ($\mathcal{O}(1'')$). On the other hand, unlensed signals will generally have non-overlapping skymaps.

\subsubsection{Data Preparation}\label{sec:data_prep}

In order to train, optimize, and test our machine learning models, we simulate the  lensed and unlensed GW signals, and inject them in Gaussian noise. Our events consist of non-spinning binary black hole mergers detectable by the LIGO-Virgo network at design sensitivity, where detectability is defined by setting a threshold of $8$ on the network SNR. 

We follow \cite{haris2018} to generate a set of strongly lensed pairs of GW events, where the source BBH mergers follow a well-motivated distribution of masses and redshifts, and the lenses are assumed to be galaxies that can be modelled as singular isothermal ellipsoids whose parameters are drawn from the SDSS galaxy population catalog \cite{sdss}. We generated $\approx  2800$ detectable lensed event pairs and $\approx  1000$  unrelated events, which corresponds to half a million unlensed pairs. We subdivide this set into two sets; we use one for training, and the other for validation. For testing, we use another, distinct, set, although the general presciption still follows \cite{haris2018} \footnote{This data set is chosen for testing because the posterior overlap statistic was already evaluated for the candidate pairs in this set (and reported in \cite{haris2018}), which allows for a ready comparison with the ML statistic.}. This set consists of $ \approx 300$ lensed pairs and $\approx 1000$ unrelated events (half a million unlensed pairs). From here on out, we refer to the training and validation data set as ``DSTrV'', and the testing set as ``DSTe''.

The waveforms are generated using the approximant $\textsc{IMRPhenomPv2}$ \cite{imrphenompv2-1, imrphenompv2-2, imrphenompv2-3}, as implemented in the $\textsc{LALSimulation}$ module of the $\textsc{LALSuite}$ software package \cite{lalsuite}. The waveforms are then projected onto the LIGO and Virgo detectors using their antenna pattern functions, as implemented in the $\textsc{pycbc}$ \cite{pycbc_software} software package. 

The detector noise is assumed to be Gaussian, and is generated using the zero-detuned high-power PSDs of Advanced LIGO and Advanced Virgo at their design sensitivities \cite{AvirgoSensitivity,aLIGOSensitivity}, as implemented in $\textsc{pycbc}$. The projected waveforms are then added to the detector noise strain to produce the total detector strain time series.

From the time series' surrounding each GW event, we generate Q-transform images for each detector. For events whose primary mass $m_1 > 60 M_{\odot}$, we set the range of quality factors to $(3,7)$; otherwise, we set the range to $(4,10)$. Further, using the same time series', we use Bayestar to generate the localisation skymaps for all the events.


\subsubsection{Feature Construction}\label{sec:features}

Comparing the shapes of two time-frequency maps can be interpreted as a problem of image recognition, and therefore lends itself nicely to a machine-learning (ML) analysis designed for such problems. Motivated by the fact that the Q-transform based time frequency maps of lensed pairs will have similar shapes (though different signal energies across time-frequency tiles), while unlensed pairs while have dissimilar shapes in general, we superimpose the time-frequency maps of candidate pairs by aligning them along the time axis, which we pass to our machine learning algorithm.

On the other hand, while lensed pairs will have overlapping skymaps and unlensed pairs will not, the \emph{shapes} of these maps are not in general expected to be the same, since the relative position of the two images with respect to the detectors are, in general, different (due to the rotation of the earth). However, GW events' localisation skymaps are probability density functions in the space of right-ascension ($\alpha$) and declination ($\delta$). Thus, a skymap can be thought of as a two-dimensional matrix where each element gives the probability density evaluated at a given pixel in the skymap's image grid spanning the space of $(\alpha, \delta)$. The products of simple operations involving the matrices of candidate pairs can then be used as features that  ML algorithms can employ to identify lensed events. 

The Bayestar localisation skymaps are usually generated in .fits format, which contains the skylocalisation posterior information sampled over an adaptive \textsc{HEALPix} grid \cite{healpix-grid}. We project them to cartesian coordinates using the \textsc{HEALPY} python library \cite{healpy-1, healpy-2}, which gives us the localisation posterior evaluated over a $400 \times 800$ rectangular grid of pixels corresponding to $(\alpha , \delta)$ pairs. Denoting the skylocalisation posteriors of each of events pertaining to a candidate lensed pair as $ P^1_{ij} = P(\alpha_i,  \delta_j~|~d_1)$ and $P^2_{ij} = P(\alpha_i , \delta_j~|~d_2)$, we can construct the following metrics which can serve as features using which we can train an ML algorithm:

\begin{equation}
\begin{aligned}
k_1 = \sum_{i}\sum_{j} P^1_{ij}P^2_{ij},  \qquad k_2 = \sum_{i}\sum_{j} |P^1_{ij}-P^2_{ij}|  \\
\qquad k_3 = \sqrt{\langle \left (P^1_{ij}P^2_{ij}) \right)^2 \rangle - \langle k_1 \rangle^2} 
\end{aligned}
\end{equation}
$k_1$ is motivated by the posterior overlap statistic \cite{haris2018}, $k_2$ is the absolute difference between the elments of the matrices, while $k_3$ is a standard deviation-like metric of the overlap between the skymaps. Note that angular brackets signify averaging over the total number of elements in each matrix.


\subsubsection{Overall Flow}

For simplicity, we build two sets of ML models - one that learns from Q-transforms and another that is fed with skymaps - to classify the event pairs as either lensed and unlensed. The models employ two different ML algorithms -- \textsc{DenseNet201} \cite{densenet} and \textsc{XGBoost} \cite{xgboost} (see Sec.~\ref{sec:ML}).

The first set consists of three \textsc{DenseNet201} ML models trained on superimposed QT (Q-Transform) images of the event pairs for each of the three detectors: H1 (Hanford), L1 (Livingston) and V1 (Virgo), operating at their design sensitivities. We further construct an \textsc{XGBoost} model trained on the output of the \textsc{DenseNet201} models. The output of this \textsc{XGBoost} model gives us the probability of the lensing hypothesis, given the Q-transform images: $P(\mathcal{H}_{L}|\text{ QT1, QT2})$ \footnote{A more complete notation for this probability would be as follows: $P(\mathcal{H}_{L}|\lbrace\text{ QT1-H1, QT2-H1}\rbrace; \lbrace\text{ QT1-L1, QT2-L1}\rbrace; \lbrace\text{ QT1-V1, QT2-V1}\rbrace)$. However, for notational simplicity, we omit the reference to the interferometers.}.

We construct another \textsc{XGBoost} model trained on the metrics derived from pairs of lensed and unlensed Bayestar skymaps. The output of this \textsc{XGBoost} model gives us the probability of the lensing hypothesis, given the Bayestar skymaps: $P(\mathcal{H}_{L}|\text{SM1, SM2})$. 

The final output of our ML classifier is then given by:
\begin{widetext}
\begin{equation}\label{eq:rankstat}
P(\mathcal{H}_{L}|\lbrace \text{QT1, QT2} \rbrace; \lbrace \text{SM1, SM2} \rbrace) = P(\mathcal{H}_{L}| \text{ QT1, QT2})\cdot P(\mathcal{H}_{L}|\text{SM1, SM2})
\end{equation}
\end{widetext}

\begin{figure}
\includegraphics[trim={0.cm 0.5cm 7.cm 0.cm},clip,width=0.98\linewidth]{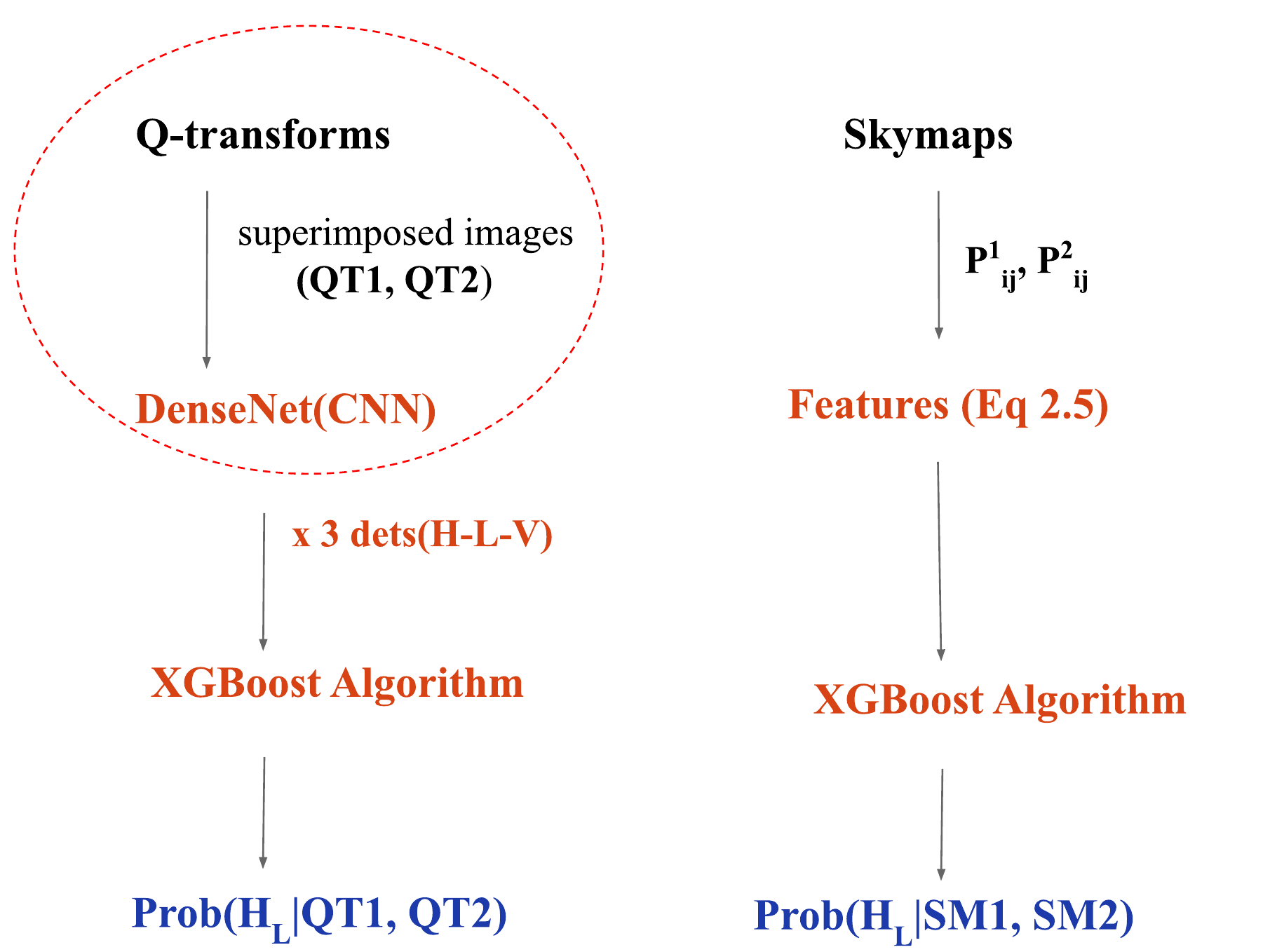}
\caption{A visual representation of the overall flow of our ML classification scheme. Note that, in principle, one could have avoided the step that trains a second \textsc{XGBoost} algorithm on features derived exclusively from the skymaps, and instead just used one \textsc{XGBoost} that jointly trains on features from the skymaps and the outputs of the \textsc{DenseNet} algorithms. We found that both methods give similar results. We therefore choose to include the additional \textsc{XGBoost} because it facilitates a stepwise analysis of the outputs of the individual components of the overall flow, trained separately on intrinsic and extrinsic parameters of the candidate pairs.}
\label{fig:flow}
\end{figure}

We summarize the overall flow of our classification scheme in Fig.~\ref{fig:flow}. 

\subsubsection{Machine Learning Models}\label{sec:ML}

In this subsection, we briefly summarize the ML algorithms we use: \textsc{DenseNet201} and \textsc{XGBoost}.

\paragraph{\textsc{DenseNet201}:} A number of supervised machine learning algorithms exist for binary classification problems. However, only a relatively small subset of these are particularly suited for image recognition. Among them is the \textsc{DenseNet} ML \cite{densenet} algorithm, which is a kind of convolutional neural network (CNN) with important improvements to mitigate problems that typically plague CNNs. A CNN, in turn, is a category of artificial neural networks (see, e.g, \cite{nn}) often used for classification problems that involve images, image recognition and computer vision (see, for example, \cite{cnn}).

The basic architecture of a neural network consists of input/output layers of neurons, and a set of hidden layers in between \cite{hastie2009}. Each neuron holds a number between in the range $[0, 1]$. An image passed to a neural network would fill the neurons of the input layer with values corresponding to the pixels of the image grid. The classification prediction of the neural network is recorded in the neurons of the output layer; specifically, in a binary classification problem such as ours, the output layer has one neuron representing the probability that the pair of superimposed Q-transforms corresponds to the ``lensed'' case.

The neurons in each hidden layer are derived using a two step process. The first step involves a linear operation between the vector of neurons $\vec{a}$ in the previous layer, and a matrix of weights $\overleftrightarrow{W}$, and the second a non-linear operation that maps the output of the linear operation to numbers in the range $[0, 1]$:

\begin{equation}\label{eq:activation}
\vec{a}_{n+1} =    f(\overleftrightarrow{W}_n \cdot \vec{a}_n + \vec{b}_n)
\end{equation}

Here, the non-linear function $f$ is referred to as the ``activation function''; common choices include the ``sigmoid function'' and ``the rectified linear unit'' (ReLu) function (see, e.g., \cite{activation-function}). Further, the vector $\vec{b}$ is called the ``bias''. This process is applied iteratively until the output layer is filled. 

Training the neural network ultimately comes down to determining an optimal choice of weights matrices and bias vectors. This can be achieved by feeding the neural network with labelled data, and penalizing the network's incorrect predictions using an appropriately defined cost function. The popular choice of cost function for binary classification is the binary cross entropy:

\begin{equation}\label{bce}
L_{\mathrm{CE}} = - [ y \log(p) + (1-y)\log(1-p) ]
\end{equation}

where $y$ is the ground-truth (``lensed'' = 1 or ``unlensed'' = 0) of the labelled data, and $p$ is the neural-network's predicted value for a given choice of weights and biases. Minimizing the loss function averaged over multiple training instances with distinct labelled data, using gradient descent, provides the required weights and biases.

In CNNs, some of the hidden layers perform convolution operations between the previous layer, and appropriately chosen filters, in place of the operation described in Eq.~\eqref{eq:activation}. The filter can be thought of as a matrix whose size is usually smaller than the matrix of pixels input to the CNN. The convolution operation then involves ``sliding'' the filter across the pixel grid matrix, which mathematically amounts to taking the product of the filter with each of the submatrices of the pixel grid matrix. The resulting output is sometimes referred to as a ``feature map''.

A \textsc{DenseNet} is a type of deep CNN.  In addition, its architecture has a few modifications to alleviate some of the problems commonly faced when using CNNs. \textsc{DenseNet}'s are based on the observation that CNNs can be substantially deeper, more accurate, and computationally efficient to train if there are shorter connections between the layers close to the input and those close to the output. Thus, in a typical \textsc{DenseNet} model, for each layer, the feature maps of all preceding layers are used as inputs. Furthermore, the current layer's own feature map is used as input to all the subsequent layers. Because of this type of architecture, \textsc{DenseNet} models have several advantages compared to other CNN models. They greatly reduce the number of parameters that define the architecture of the neural network, mitigate the vanishing-gradient problem, encourage feature reuse and strengthen the feature propagation through the network.

\paragraph{\textsc{XGBoost}:} eXtreme Gradient Boosting, (\textsc{XGBoost}) \cite{xgboost} is a type of ensemble classifier that uses the combined output of a collection of trained decision trees to provide a probabilistic prediction of class-membership to data that needs to be seggregated into discrete categories. A decision tree, in turn, learns from training data by iteratively placing linear cuts in feature-space which minimizes an appropriately chosen loss function. The repeated splits result in the seggregated data being pushed down two separate branches at each leaf node in the tree, starting from the root-node where the first split in the training data takes place, and ending at leaf nodes where a terminating criterion (e.g: minimum number of samples in a leaf) has been satisfied.

``Bagging'' (see, e.g., \cite{bagging}) and ``boosting'' (see, e.g, \cite{boosting}) are two ways in which the outputs of decision trees can be combined. In bagging, bootstrapped copies of the training data are passed to a collection of decision trees. The trees are then fitted, in parallel, to the training data they receive, and the final prediction of the classifier is an average over all the outputs across the ensemble of trees \cite{randomforest}. In contrast, boosting algorithms such as \textsc{XGBoost}, fit decision trees to training data sequentially, where each subsequent tree improves on the errors in the predictions of class probability of the preceding tree. 

In eXtreme Gradient Boosting, the iterative process of incrementally improving the prediction of the classifier with every fitted decision tree, reduces to minimizing the following objective function \cite{xgboost}:

\begin{equation}
\mathcal{L}^{\mathrm{obj}}_{t+1} = \sum_i \mathcal{L}(y^i, p^{i}_{t+1} = p^{i}_{t} + O_{t}) + \gamma T + \frac{1}{2}\lambda O^2_t
\end{equation}

where, as before, $y^i$ is the ground truth of training data point $i$, $p^{i}_{t} (p^{i}_{t+1})$ is the classifier's predicted probability of class membership after the sequential fitting of $t$ ($t+1$) trees. For binary classification problems such as the one we are trying to tackle, the loss function $\mathcal{L}$ is simply the binary cross-entropy defined in Eq.~\eqref{bce} (summed over the entire training set), and $O_{t}$ is the output of the decision tree $t$ with respect to which the objective function is to be minimized. The piece $ \gamma T + \frac{1}{2}\lambda O^2_t$ in the objective function is a regularization term that controls the classifier's tendency towards overfitting by reducing its sensitivity to individual training data points. Here, $T$ is the total number of leaves in a tree, and $\lambda, \gamma$ are hyperparameters that can be appropriately set depending on the data at hand.

Minimizing $\mathcal{L}^{\mathrm{obj}}$ for each decision tree (which can have a vast variety of structures) is in general highly complicated. \textsc{XGBoost} thus simplifies the minimization process in two ways. The first is that the loss function is approximated by a second-degree Taylor polynomial in $O_t$. The second is that within each tree, the objective function is repeatedly minimized at each leaf node. As a result, the process of fitting a decision tree reduces to maximizing the gain when splitting the training data at each leaf node. The gain is defined as the difference between the sum of the similarity scores of the two daughter nodes post the split, and the similarity score of the parent leaf. The similiarity score at a leaf node $l$ (containing $N_l$ training samples) in tree $t+1$ is defined as \cite{xgboost}:
\begin{equation}
S^l_{t+1} = \frac{(\sum_i^{N_l} R^i_t)^2}{\sum_i p^i_t(1-p^i_t) + \lambda}
\end{equation}
where the sum is taken over all the samples in the leaf node, and $R^i_{t} \equiv p^i_t - y^i$ is the residual of the $i$th training data point in the leaf node. The output of each tree, defined as $S^{term}_{t+1}/\sum_i R^i_t$ for the terminal leaf node, is then rescaled by a user defined learning rate $\eta$ and then added to the log of the odds ratio corresponding to $p^i_t$, from which the probability estimate of tree $t+1$ can be trivially computed.

As mentioned earlier, $\lambda, \gamma$ are user defined regularization parameters that control overfitting. Specifically, $\gamma$ sets a threshold on the gain; leaves along branches whose gains do not exceed $\gamma$ are pruned. Thus, since positive values of $\lambda$ tend to reduce the gain, $\lambda$ effectively encourages pruning, which in turn reduces the sensitivity of the decision tree to individual training data points. \footnote{In ML literature, $\lambda$ is often referred to as a ``regularization parameter'' and $\gamma$ is referred to as a ``tree complexity parameter''.}

\subsubsection{Training and optimisation}

\paragraph{\textsc{DenseNet201}}: We use a \textsc{DenseNet} pretrained on the ``Imagenet dataset'' \cite{imagenet}, which allows it to pick up features common to most images. We then add fully connected layers to it, along with the final layer of just one neuron, for our binary classification, and then retrain it with data specific to our problem (to wit, the superimposed Q-transforms).  This method of pretraining with a generic data set and then retraining with a more specific one, is called ``transfer learning". The most significant benefit of this method is that it reduces the size of the dataset required for training and solving the problem at hand.

For each of the three detectors  H1, L1 and V1, we train three individual \textsc{DenseNet201} models using superimposed Q-transform pairs, where each image corresponds to a 3-dimensional array ($128 \times 128 \times 3$) of pixels \footnote{Each pixel contains RGB values that correspond to the normalized signal energy at discrete time-frequency coordinates in the Q-transform image.}. The \textsc{DenseNet} model is loaded with the imagenet weights using the neural network
package \cite{keras}. To make it suitable for our binary classification task, its top layer is removed and a dense layer of $256$ neurons with the ReLu activation function is added along with the final output layer of a single neuron with a sigmoid activation function. Each of the three models is trained on an equal number ($1400$) of lensed and unlensed Q -transform image pairs subselected from the DSTrV dataset using TPU (Tensor Processing Unit) hardware, which is available in a \textsc{kaggle} notebook \cite{kaggle}. In the top fully connected layer of the network, we use the sigmoid activation function (see. e.g., \cite{sigmoid}) and we employ the Adam optimizer \cite{adam} for efficient gradient calculations. 
The model prediction is validated using a validation set sub-selected from the total training set. 



\paragraph{\textsc{XGBoost}}: As described in the previous section, \textsc{XGBoost} has a number of tunable hyperparameters that need to be set based on the problem at hand. 

The hyperparameter ``n$\_$estimators'' sets the number of decision trees in the ensemble classifier that are to be fit to the training data sequentially. It can equivalently be thought of as the number of fitting iterations the model goes through as it sequentially improves the prediction of the ensemble classifier. We set n$\_$estimators to $110$. The learning rate, regularization parameter and tree complexity parameter are set to their default values of $0.3, 1, 0$ respectively. The maximum depth of each decision tree is set using $\text{max\_depth} = 6$.

In addition, we also set the ``scale$\_$pos$\_$weight'' parameters to $0.01$. This hyperparameter serves as a weight to account for training data being biased towards one class -- in our case, the unlensed class, for which we had about $100$ times more data points than for the lensed class.

The first \textsc{XGBoost} model is trained on the features derived from lensed and unlensed pairs of skymaps, described in Sec.~\ref{sec:features}, using the ``DSTrV'' dataset. Additionally, a second \textsc{XGBoost} model is trained on the outputs of each of the three \textsc{DenseNet} models. The outputs of the two \textsc{XGBoost} models are then combined (cf.~\ref{eq:rankstat}) to provide a ranking statistic for candidate lensed pairs.
 

\section{Results}\label{sec:results}
	\subsubsection{Testing and Cross-validation}
We assess the performance of the trained ML models on the ``DSTe'' dataset. This allows us to compare their performance with the posterior overlap statistic, which is already computed for this dataset \cite{haris2018}. We summarize the performance of the ML models and the posterior overlap statistic with ROC \footnote{Receiver Operating Characteristic} plots of efficiency vs false positive probability (FPP), where efficiency is the ratio of accurately classified lensed events to the total number of lensed events, and FPP is the ratio of wrongly classified unlensed events to the total number of unlensed events. 

To check the robustness of the outputs of the machine learning models to changing training sets , we use stratified k-fold cross validation. We implement cross validation by doing a round-robin of dividing our dataset into $k = 3$ ($k = 10$) parts for the \textsc{DenseNet} (\textsc{XGBoost}) models, using one part for validation and the rest for training. We test the $k$ trained machines with the DSTe dataset. 

\subsubsection{ROC Plots}
We evaluate the performance of the overall classifier and its different components using ROCs. For comparison, we also plot the ROCs for the posterior overlap statistic. We first test the performance of the individual \textsc{DenseNet} models trained on Q-transforms pertaining to each of the three detectors: H1, L1 and V1. We then test the \textsc{XGBoost} model trained on the outputs of the \textsc{DenseNet} models. Since we used cross-validation to assess the robustness of the models, we trained and validated each of the models on the different cross-validation subsets of the DSTrV data set, and tested the differently trained models on the DSTe data set. This gives us an estimate of the variation of the ROCs due to differences in the training set.

\begin{figure}[htb]
\includegraphics[width = \linewidth]{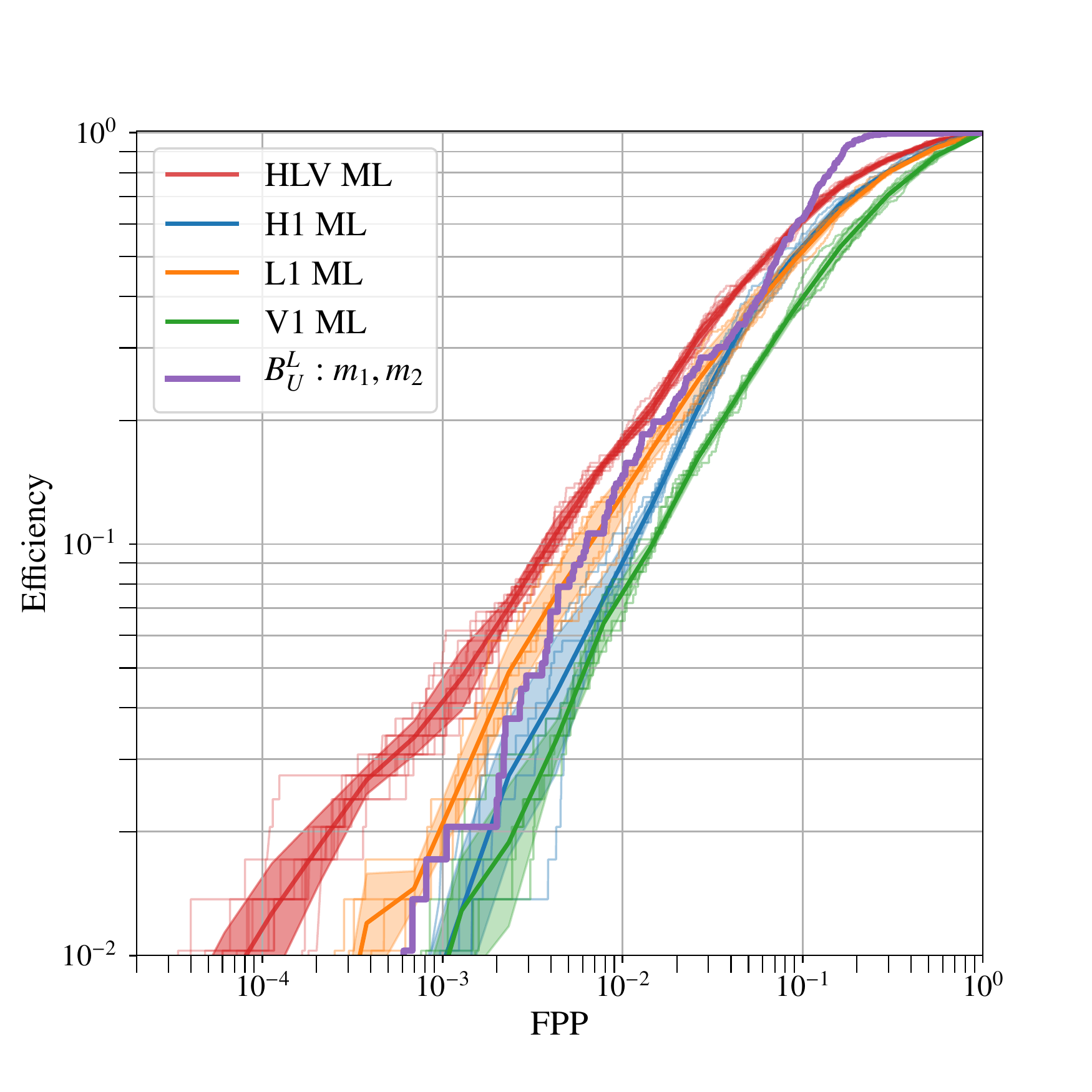}
\caption{ROCs for \textsc{DenseNet} models trained on lensed and unlensed pairs of superimposed Q-transforms, for different cross-validation subsets of the DSTrV training set. ROCs for models trained on Q-transforms corresponding to individual detectors are evaluated, in addition to ROCs pertaining to the \textsc{XGBoost} model trained on the outputs of the individual \textsc{DenseNet} models. For comparison, the ROC for the posterior overlap statistic that uses parameter estimation posteriors on the component masses, $m_1, m_2$, is also plotted. At low false positive probabilities, the individual DensetNet models perform comparably to the posterior-overlap statistic. On the other hand, the \textsc{XGBoost} model produces efficiencies that are $1.5 - 2$ times better than the posterior overlap statistic at low FPPs, although there is some variation in the ROCs when the training set is changed, caused by small-number statistics. These improvements at low FPPs must therefore be interpreted with some caution.}
\label{fig:roc_qts}
\end{figure}
Fig.~\ref{fig:roc_qts} plots ROCs for the outputs of these models trained on Q-transforms. The ROC for the posterior overlap statistic constructed using parameter estimation posteriors on the component masses ($m_1, m_2$), is also plotted for comparison. The ROCs pertaining to the individual \textsc{DenseNet} H1, L1, V1 models perform similarly to the ROC for the posterior overlap statistic, both at low and high false positive probabilities. The mean ROC corresponding to the \textsc{XGBoost} model trained on the outputs of the individual \textsc{DenseNet} models performs comparably to the posterior overlap statistic. At very low FPPs, ML seems to perform about $1.5-2$ times better than the posterior overlap statistic. However, there is some variation in the \textsc{XGBoost} model's ROC due to the changing training set. These improvements must therefore be interpreted with some caution. As the variation in the ROCs at these FPPs suggests, low-number statistics are likely causing the ROC to be sensitive to changes in the training set.

\begin{figure}[htb]
\includegraphics[width = \linewidth]{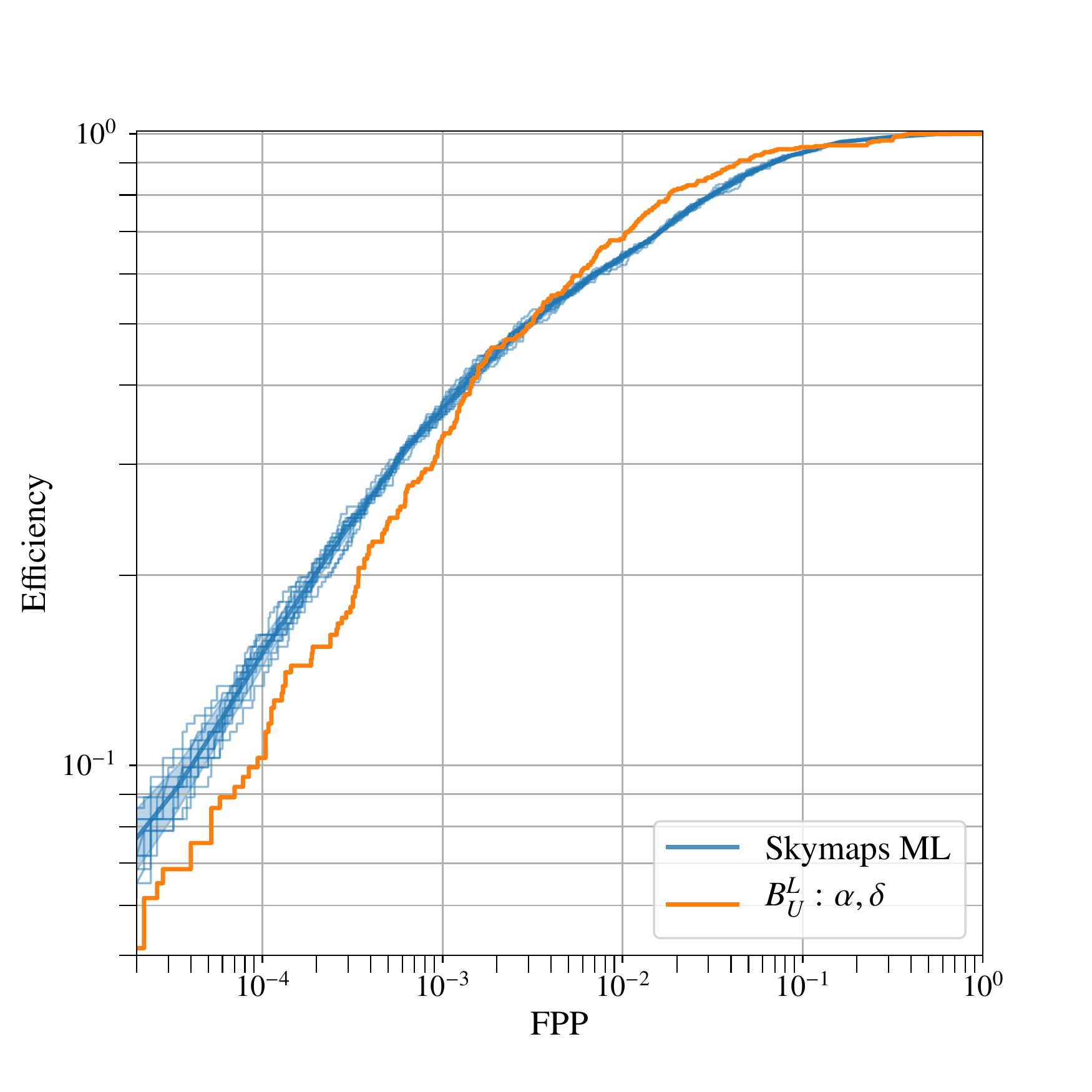}
\caption{ROCs for the \textsc{XGBoost} model trained on metrics derived from pairs of Bayestar localisation skymaps, for different cross-validation subsets of the DSTrV trainsing set. For comparison, the ROC for the posterior overlap statistic that uses parameter estimation posteriors on the skylocation coordinates, $\alpha, \delta$, is also plotted. The \textsc{XGBoost} performs almost as well (in fact, marginally better if only the mean ROC is considered), as the posterior overlap statistic, at low false positive probabilities.}
\label{fig:roc_skymaps}
\end{figure}
Fig.~\ref{fig:roc_skymaps} plots ROCs for the \textsc{XGBoost} model trained on the features (metrics) derived from pairs of Bayestar skymaps. Each ROC pertains to a different cross-validation subset of the DSTrV dataset. The ROC for the posterior overlap statistic evaluated using only the right-ascension ($\alpha$) and declination ($\delta$) is plotted for comparison. The \textsc{XGBoost} performs as well as the posterior overlap statistic at low false positive probabilities, although at higher false positive probabilities the latter performs marginally better. As with the \textsc{DenseNet} models, there is some variation in the ROCs when the training set is varied. 

\begin{figure}[htb]
\includegraphics[width = \linewidth]{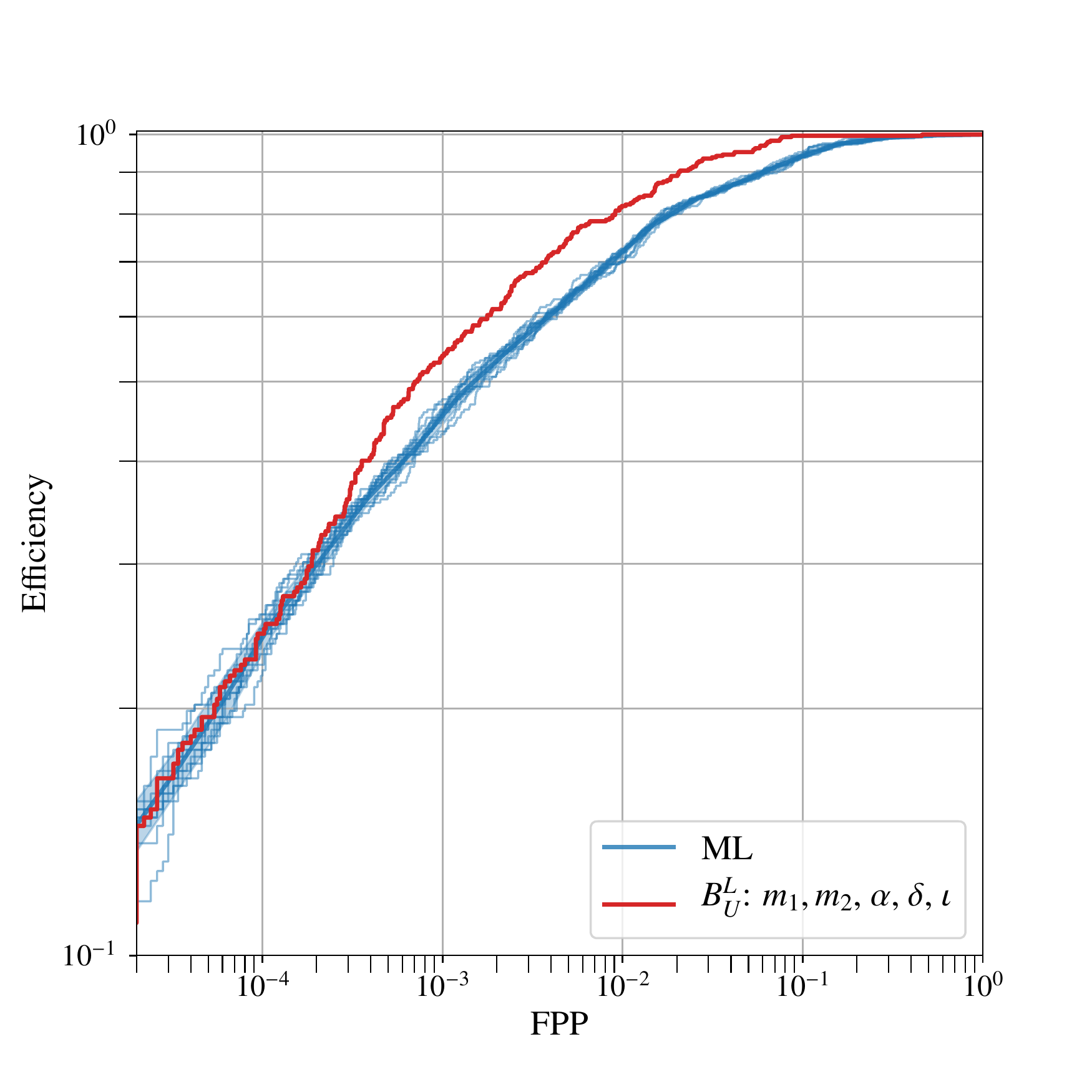}
\caption{ROCs for the overall classifier, for different cross-validation subsets of the DSTrV trainsing set. Note that the output of the overall classifier is the output of the \textsc{XGBoost} model trained on the ouputs of the three \textsc{DenseNet} models pertaining to H1, L1 and V1, as well as the ouput of the first \textsc{XGBoost} model trained on Bayestar skymaps. At low false positive probabilities, the classifier performs almost identically to the posterior overlap statistic, with mild variation in the ROCs when the training data set is varied.}
\label{fig:roc_combined}
\end{figure}
Fig.~\ref{fig:roc_combined} plots ROCs for the overall classifier, which is an \textsc{XGBoost} model trained on the outputs of the \textsc{DenseNet} models and the first \textsc{XGBoost} model. For comparison, the ROC for the posterior overlap statistic evaluated using the parameter estimation posterior on $m_1, m_2, \alpha, \delta$ is also plotted. The mean ROC for the overall classifier performs almost identically to the posterior overlap statistic at low false positive probabilities, although at higher false positive probabilities the posterior overlap statistic performs marginally better.

\section{Summary and Outlook}\label{sec:conclusion}
	GW observations of BBH events is expected to increase significantly in future observing runs, with $\mathcal{O}(10^2)$ events during O4 and $\mathcal{O}(10^5 - 10^6)$ during the 3G era. The number of candidate lensed pairs to classifiy could therefore be as high as $\mathcal{O}(10^4)$ and $\mathcal{O}(10^{10} - 10^{12})$, respectively. Current optimal Bayesian methods, such as the posterior overlap statistic, rely on the parameter estimation posterior on the source parameters, which could take anywhere from several hours to several days to sample. 

This therefore motivates the need to come up with a preliminary classification scheme, that can rapidly rule out the vast majority of unlensed candidates. To that end, as a proof-of-principle, we construct a machine learning based classifier that can classify pairs of non-spinning BBH events in seconds. We use two ML algorithms: DenseNet201 and XGBoost, to build models trained on time frequency maps and Bayestar skymaps of pairs of events. We construct 3 DenseNet models trained on GW events projected onto each of the three detectors in the LIGO-Virgo network at design sensitivity. The outputs of these models are fed to an XGBoost classifier to construct a corresponding model. The output of this model is then combined with the ouput of another XGBoost model trained on pairs of lensed and unlensed Bayestar skymaps, to produce the final ranking statistic of our overall ML classifier (cf. Fig.~\ref{fig:flow} and Eq.~\ref{eq:rankstat}).

We train and validate the classifier on cross-validation subsets of the DSTrV dataset, and test the performance of the ML classifier (including its different components) on the DSTe dataset. We find that the overall ML classifier performs comparably to the posterior overlap statistic evaluated from the parameter estimation posterior on $m_1, m_2, \alpha, \delta$. More specifically, the performance of the ML classifier, as captured by ROC plots, shows that at low false positive probabilities, the classifier performs almost identically to the posterior overlap statistic, although at high false positive probabilities, the performance of the latter is marginally better. 

Simple benchmarking tests suggest that our trained ML classifier is able to classify each event within $2-3$ seconds \footnote{Note that this time is largely taken up in loading the necessary files for classification. The classification step itself takes less than a second.}. Including the time to produce the Q-transform images and Bayestar skymaps, the total classification time is still less than a minute. This is significantly faster than the posterior overlap statistic, which takes several minutes to classify once the parameter estimation posteriors are available. Since, in addition, these posteriors themselves can take hours to days to produce, per event, the benefit of using ML to perform a preliminary sweep of lensed candidate pairs to rule out the vast majority of them as unlensed, becomes manifestly evident. 

Additionally, rapid ranking of candidate pairs makes estimating a background distribution computationally feasible.  Such a distribution enables assigning statistics such as p-values/false positive probabilities, which are often the preferred statistics since they can be interpreted independently of the models used to analyze the pairs. Another potentially useful application of the rapid identification (and dissemination) of lensed GW events is in multi-messenger astronomy, since the joint GW-EM detection of lensed events could enable important tests of general relativity.

It might be worth mentioning that in addition to the posterior overlap statistic, there are more comprehensive Bayesian classification methods that take even longer to run. A fully Bayesian, joint parameter estimation scheme to identify lensed pairs by evaluating a coherence ratio that accounts for correlations between parameters of lensed events, and selection effects, currently takes of the order of weeks to complete, per candidate pair \cite{hanabi, Liu2021}. A more approximate joint parameter estimation method that neglects selection effects, is found to identify lensed pairs with similar efficiencies as the full joint parameter method, but within hours instead of weeks \cite{golum}. Thus, identifying lensed pairs from the enormous number of candidate pairs in future observing runs, can follow a step-wise procedure, where an ML classification method such as ours can rapidly rule most of the candidate pairs as unlensed. The surviving pairs can then be followed up by the posterior overlap statistic, and then by joint parameter estimation methods.

Note that our work assumed stationary Gaussian noise, and that the candidate pairs consist of confirmed, high-significance non-spinning BBH events. We plan to systematically relax these assumptions in future work. Specifically, we are currently looking at the possibility of classifying confident GW events in real noise. We plan to train the machine on events injected in real noise, whiten the data so that the Q-transforms are less sensitive to varying PSDs, and investigate the possibility of using additional features. We are also working towards the classification of marginal BBH events, with an ML scheme similar to what was presented in this work. 
We hope to report the results of these investigations in the near future.

\section{Acknowledgements} \label{sec:ack}

We would like to thank Deep Chatterjee and Otto Hanuksela for useful discussions. We thank K. Haris for providing the DSTe dataset used in \cite{haris2018}. We also thank Pinak Mandal and Shashank Roy for helping SG learn various machine learning techniques. SG's, SJK’s, and PA’s research was supported by the Department of Atomic Energy, Government of India. In addition, SJK’s research was supported by
the Simons Foundation through a Targeted Grant for the Mathematical and Physical Sciences to ICTS-TIFR. PA’s research was supported by the Max Planck Society through a Max Planck Partner Group at ICTS and by the Canadian
Institute for Advanced Research through the CIFAR Azrieli Global Scholars program. This research has made use of data, software and/or web tools obtained from the Gravitational Wave Open Science Center (https://www.gwopenscience.org), a service of LIGO Laboratory, the LIGO Scientific Collaboration and the Virgo Collaboration. LIGO is funded by the U.S. National Science Foundation. Virgo is funded by the French Centre National de Recherche Scientifique (CNRS), the Italian Istituto Nazionale della Fisica Nucleare (INFN) and the Dutch Nikhef, with contributions by Polish and Hungarian institutes. Some of the computations were performed with the aid of the Alice computing cluster at ICTS-TIFR, and rest on \textsc{kaggle} \cite{kaggle} with TPU hardware acceleration.

\bibliography{references}


\end{document}